\documentclass[fp,twocolumn]{jpsj3}
\usepackage{txfonts}
\usepackage{graphicx}
\usepackage{amssymb, amsfonts, amsmath}
\usepackage{dcolumn}
\usepackage{bm}
\usepackage{color}
\usepackage{mathrsfs}
\usepackage{lineno}
\usepackage[normalem]{ulem}
\definecolor{lblue}{RGB}{102,102,225}

\emergencystretch=\maxdimen
\newlength{\figwidth}
\newcommand{\fig}[3]

\title{Discovery of Emergent Photon and Monopoles in a Quantum Spin Liquid}

\author{Y.\;Tokiwa$^{1,2}$, T.\;Yamashita$^1$, D.\;Terazawa$^1$, K.\,Kimura$^{3,4}$, Y.\;Kasahara$^1$, T.\;Onishi$^1$, Y.\;Kato$^5$, M.\;Halim$^4$, P.\;Gegenwart$^2$, T.\;Shibauchi$^6$, S.\;Nakatsuji$^{4,7}$, E.-G.\;Moon$^8$, Y.\;Matsuda$^1$}

\inst{$^1$Department of Physics, Kyoto University, Kyoto 606-8502, Japan \\
$^2$Center for Electronic Correlations and Magnetism, Institute of Physics, Augsburg University, 86159 Augsburg, Germany \\
$^3$Division of Materials Physics, Graduate School of Engineering Science, Osaka University, Toyonaka, Osaka 560-8531, Japan \\
$^4$Institute for Solid State Physics, University of Tokyo, Kashiwa, Chiba 277-8581, Japan \\
$^5$Department of Applied Physics, University of Tokyo, Bunkyo, Tokyo 113-8656, Japan \\
$^6$Department of Advanced Materials Science, University of Tokyo, Kashiwa, Chiba 277-8561, Japan \\
$^7$CREST, Japan Science and Technology Agency, Kawaguchi, Saitama 332-0012, Japan \\
$^8$Department of Physics, Korea Advanced Institute of Science and Technology, Daejeon 305-701, Korea} 

\abst{Quantum spin liquid (QSL) is an exotic quantum phase of matter whose ground state is quantum-mechanically entangled without any magnetic ordering. A central issue concerns emergent excitations that characterize QSLs, which are hypothetically associated with quasiparticle fractionalization and topological order. Here we report highly unusual heat conduction generated by the spin degrees of freedom in a QSL state of the pyrochlore magnet Pr$_2$Zr$_2$O$_7$, which hosts spin-ice correlations with strong quantum fluctuations.  The thermal conductivity in high temperature regime exhibits a two-gap behavior, which is consistent with the gapped excitations of magnetic ($M$-) and electric monopoles ($E$-particles).  At very low temperatures below 200\,mK, the thermal conductivity unexpectedly shows a dramatic enhancement, which well exceeds purely phononic conductivity, demonstrating the presence of highly mobile spin excitations. This new type of excitations can be attributed to  emergent photons ($\nu$-particle), coherent gapless spin excitations in a spin-ice manifold. }

\begin{document}
\maketitle

\section{Introduction}

Quantum spin liquid (QSL) states show extreme quantum entanglement that is manifested by exotic elementary excitations such as topological defects and gauge fluctuations~\cite{Wen2002,Sachdev,Balents}. Despite tremendous efforts in the past years, it has remained a great challenge to identify the elementary excitations, which is a key to understanding QSL. A rare earth pyrochlore oxide, where magnetic ions form an array of corner-sharing tetrahedra, is one of the perfect venues to host QSLs. The magnetic moments exhibit strong geometrical frustration, producing a range of exotic magnetic excitations. For classical Ising spins with a strong easy-axis anisotropy, frustration results in macroscopically degenerate spin-ice states. A spin-flip from the classical spin-ice manifold can be described as a pair creation of magnetic monopoles, which interact with each other through the static Coulomb interaction~\cite{Castelnovo-Nature08,Jaubert,Morris-Science09,Ramirez-Nature99,Bramwell-Science01,Bramwell-PRL01,Fennell-Science09}. Additional spin interactions such as a transverse exchange term endow the spins with quantum fluctuations, which may lead to a QSL state by lifting macroscopic degeneracy~\cite{Hermele-PRB04,Onoda,Benton-PRB12,Machida, Savary-PRL12,Ross-PRX11,Shannon-PRL12,Gingras-RPP14,Kato-PRL15}. Although the emergence of exotic quasiparticle excitations out of a QSL state in pyrochlore quantum magnets has been widely discussed, their identifications remain largely unexplored. 

Pr$_2$Zr$_2$O$_7$ is one of the most promising candidates to observe the exotic excitations~\cite{Kimura}. Specific heat and magnetic susceptibility measurements on different samples indicate absence of magnetic order down to $\sim$30\,mK~\cite{Kimura,Matsuhira,Petit,Kimura14}, in contrast to Yb$_2$Ti$_2$O$_7$ whose ground state is recently reported to develop ferromagnetism at 200\,mK~\cite{Chang,Tokiwa}. The neutron scattering experiments have reported the presence of spin-ice correlations with strong quantum fluctuations~\cite{Kimura}, different from classical spin ices such as Dy$_2$Ti$_2$O$_7$ and Ho$_2$Ti$_2$O$_7$~\cite{Gardner-RMP10,Gingras-RPP14,Sondhi12}.  

To reveal the quasiparticle excitations generated by spin degrees of freedom in this compound,  the thermal conductivity $\kappa$ is a particularly suitable probe because of the following reasons. Thermal conductivity is a highly sensitive probe for itinerant excitations and is not affected by localized degrees of freedom  in sharp contrast to specific heat, in which large nuclear Schottky contributions are observed.  In fact,  low-energy exotic spin excitations in some QSL materials have been clearly detected by the thermal conductivity~\cite{Tokiwa,SaitoPRE96,Sologubenko-PRB01,Hess-PRL04,Yamashita-NatP,Yamashita-Science}. Moreover, since the characteristic energy scale of the magnetic interactions of Pr$_2$Zr$_2$O$_7$ is a few hundred $\mu$eV,   the low-energy coherent magnetic excitations are expected to have  orders of magnitude  smaller energy scale.   It is therefore hard to detect such low-energy ($\sim$$\mu$eV) excitations with conventional neutron scattering experiments.
   
Here we report on measurements of thermal conductivity in Pr$_2$Zr$_2$O$_7$. We show that temperature dependence of thermal conductivity below 6\,K exhibits a characteristic three-step behavior, demonstrating the presence of exotic spin excitations inherent in the QSL state of Pr$_2$Zr$_2$O$_7$. The result is highlighted by a dramatic enhancement of $\kappa/T$ below 200\,mK in zero field, which indicates the emergence of highly mobile gapless spin excitations. 

\section{Experiment}

Single crystals of Pr$_2$Zr$_2$O$_7$ were grown by the floating zone method. The X-ray diffraction measurements confirmed the high crystallinity and the single phase with the lattice constant of 10.71\,\AA. Stoichiometry of the samples was ensured by monitoring the sample mass through the crystal growth process. The sample is green in color, which indicates no inclusion of Pr$^{4+}$ ions. The above observation shows high-quality of the samples. We note that a sample on the same badge exhibits a metamagnetic transition between Kagome-ice and 3-in-1-out states for the field along [111] direction~\cite{future}. This clearly shows that broadening of energy levels is significantly smaller than the magnetic exchange interaction.

The sample size is 1.8$\times$0.8$\times$0.27\,mm$^3$ for $\bm{Q}$$\parallel$$[1 \bar{1} 0]$, and 1.8$\times$1.0$\times$0.28\,mm$^3$ and 1.8$\times$0.35$\times$0.14\,mm$^3$ for sample \#1 and \#2, respectively for $\bm{Q}$$\parallel$$[111]$ and  thermal currents were applied  along the longest crystal directions. It has been reported that deviation in the composition causes variation of Curie-Weiss temperature, $\theta$~\cite{Koohpayeh}. We found only small difference in $\theta$ of sample pieces cut near to our samples ($\theta=-0.1 \sim-0.2$\, K), indicating little variation in the composition.

\section{Results and discussion}

\begin{figure}[t]
	\begin{center}
		\includegraphics[width=0.9\linewidth]{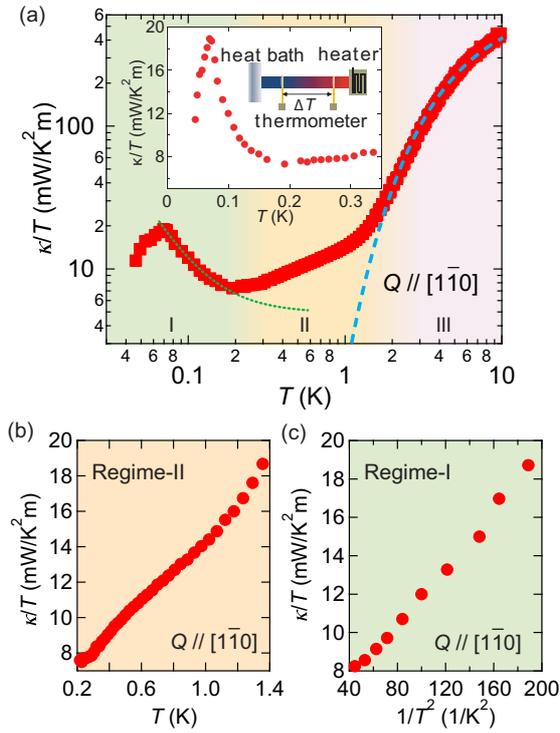}
		\caption{
		(a) Thermal conductivity divided by temperature $\kappa/T$ of Pr$_2$Zr$_2$O$_7$ as a function of temperature in zero field. Heat current $\bm{Q}$ is applied along $[1 \bar{1} 0]$ direction. Thermal conductivity shows characteristic behaviors in three different temperature regimes, I, II and III. Blue dashed line is an exponential dependence in regime-III, and green dotted line is a $1/T^2$-dependence in regime-I. 
Inset is a blowup of low-temperature region with a schematic experimental setup. 
		(b) $\kappa/T$ as a function of $T$ in regime-II. 
		(c) $\kappa/T$ plotted against $1/T^2$ in a temperature range between 70 and 160\,mK in regime-I. 
}
	\end{center}
	\vspace{-5mm}
\end{figure}

 Figure\,1(a) shows the temperature ($T$) dependence of $\kappa/T$ in zero field with the thermal current $\bm{Q}$$\parallel$$[1\bar{1}0]$. There are three temperature regimes with different $T$-dependences.  At high temperature above $T_{\rm II-III}\sim$1.5\,K (regime-III), $\kappa/T$ increases steeply with $T$. In regime-II below $T_{\rm II-III}$, $\kappa/T$ exhibits a small hump-like structure at around $\sim$0.6\,K, below which $\kappa/T$ starts to bend further downward [Fig.\,1(b)]. Upon entering regime-I below $T_{\rm I-II}\sim$200\,mK, $\kappa/T$ exhibits a striking enhancement, showing a $1/T^2$-dependence [Figs.\,1(a) and 1(c)].  After peaking at $\sim$70\,mK, $\kappa/T$ decreases rapidly with decreasing $T$.   As shown later, essentially similar $T$-dependence of $\kappa/T$ is observed for $\bm{Q}$$\parallel$$[111]$. 

\begin{figure}[t]
	\begin{center}
		\includegraphics[width=0.75\linewidth]{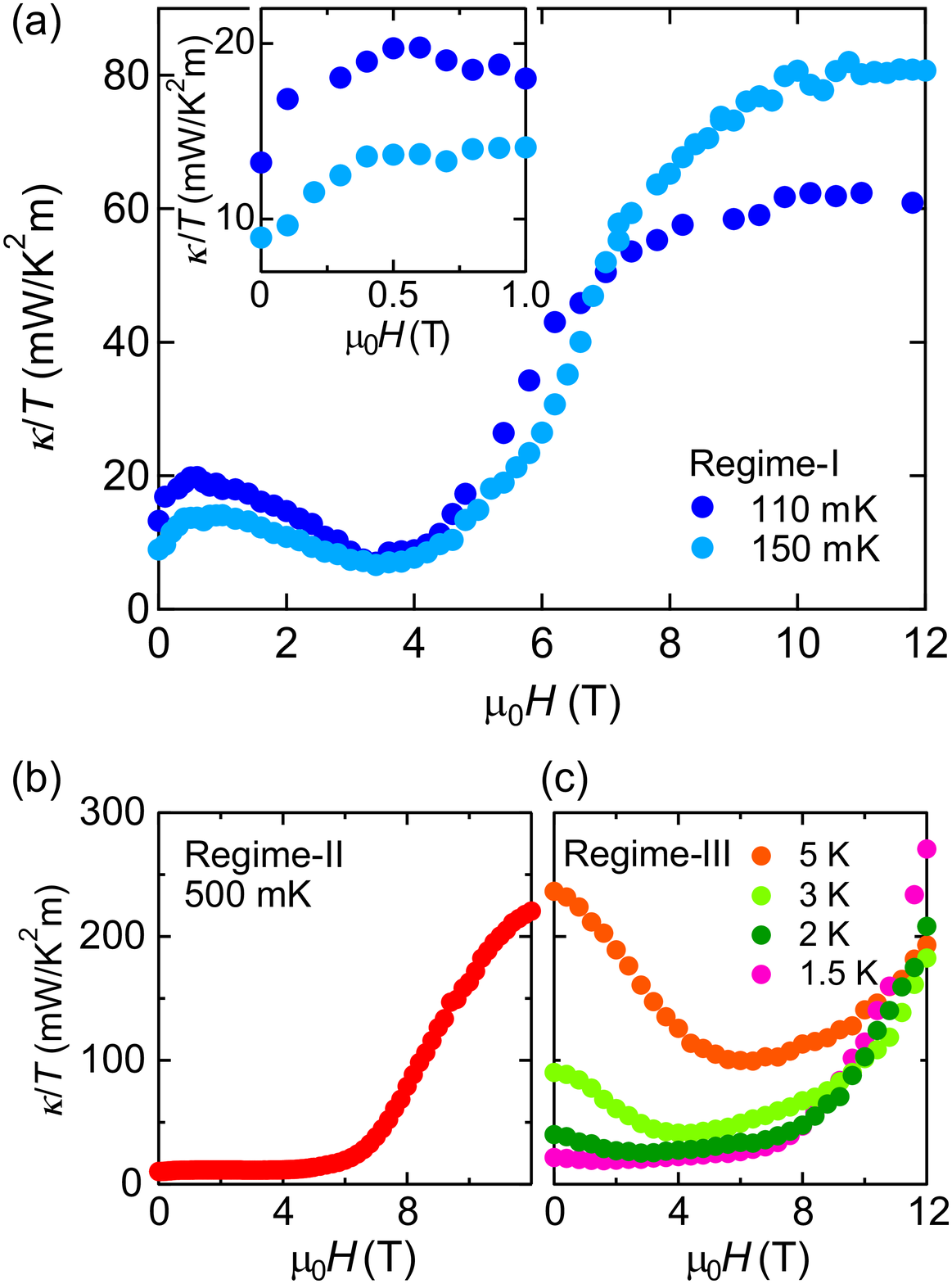}
		\caption{  
Thermal conductivity divided by temperature $\kappa/T$ as a function of magnetic field in temperature regimes (a) I, (b) II, and (c) III [see Fig.\,1(a) for the temperature regimes]. Heat current {\boldmath $Q$} is applied along $[1\bar{1}0]$ direction. Magnetic field is applied along $[111]$ direction. The inset of  (a) displays the low field behavior of $\kappa/T$.}
	\end{center}
	\vspace{-5mm}
\end{figure}

At low fields, the three different $T$-regimes reveal their own characteristic magnetic field ($H$) dependences of $\kappa(H)/T$. In regime-I, $\kappa(H)/T$ increases steeply and then decreases with $H$ [Fig.\,2(a) and its inset]. $\kappa(H)/T$ shows a nearly $H$-independent behavior in regime-II, [Fig.\,2(b)], while in regime-III, $\kappa(H)/T$ decreases initially with $H$ [Fig.\,2(c)]. On the other hand, at high fields above $\sim$4\,T, $\kappa(H)/T$ increases steeply with $H$ in all $T$-regimes. In regime-I, $\kappa(H)/T$ shows a saturation behavior at higher fields.

 Figures\,3(a) and 3(b) depict $T$-dependence of $\kappa/T$ of two different crystals (\#1 and \#2) for $\bm{Q}$$\parallel$$[111]$.    Similar to  $\bm{Q}$$\parallel$$[1\bar{1}0]$, steep increase above $\sim$1.5\,K (regime-III),  the hump structure at $\sim$500\,mK (regime-II),  and enhancement below $\sim$200\,mK (regime-I) are observed. In addition, $H$-dependence of $\kappa/T$ shown in Fig.\,3(c) is essentially similar to that for $\bm{Q}$$\parallel$$[1\bar{1}0]$ in regime-I [Fig.\,2(a)], including initial steep increase and $H$-independent $\kappa/T$ above 4\,T.  These indicate that behaviors of $\kappa/T$ are independent of $\bm{Q}$-direction.  We note that  $\kappa/T$ for \#1 and \#2 for $\bm{Q}$$\parallel$$[111]$ and that for $\bm{Q}$$\parallel$$[1\bar{1}0]$ well coincide above 1.5\,K, while  the enhancement of $\kappa/T$ below $\sim$200\,mK is sensitive to sample quality.

In Pr$_2$Zr$_2$O$_7$,  there are two main heat carriers from spin and lattice degrees of freedom. Even though Pr nuclear degrees of freedom exist as manifested in specific heat measurements, one can rule out a possibility that the nuclei carry heat because the nuclei are well localized. Let us consider the example of iso-structural classical spin ice, Ho$_2$Ti$_2$O$_7$~\cite{Toews}. It shows $H$-independent thermal conductivity at low temperatures, but its specific heat shows striking enhancement due to the level splitting of Ho nuclear spins. Therefore, it is obvious that the nuclear spin is localized contributing to the specific heat but not to the thermal conductivity. Furthermore, the non-monotonic $H$-dependence of $\kappa/T$ in regime-I (Fig.\,3a) is inconsistent with  nuclear level splitting that increases with $H$.   We also emphasize that both specific heat and magnetic susceptibility show the absence of magnetic ordering of 4$f$-moments  and  nuclear moments  down to very low temperature (30\,mK)  in Pr$_2$Zr$_2$O$_7$~\cite{Kimura}.  This excludes the magnon thermal conductivity in an ordered state. 


Remarkably, our $\kappa(H)/T$  measurements allows us to identify dominant heat carriers in regime-I. 
We note that in high field regime where $\kappa(H)/T$ is saturated, the heat conduction is mediated only by phonons. In a high enough field, where the Zeeman energy exceeds the exchange and thermal energies, spins are nearly fully polarized and gapped, and hence they cannot carry the heat. Then, spin thermal conductivity as well as spin-phonon scattering are absent, leading to $H$-independent purely phononic thermal conductivity.

\begin{figure}[t]
	\begin{center}
		\includegraphics[width=1.0\linewidth]{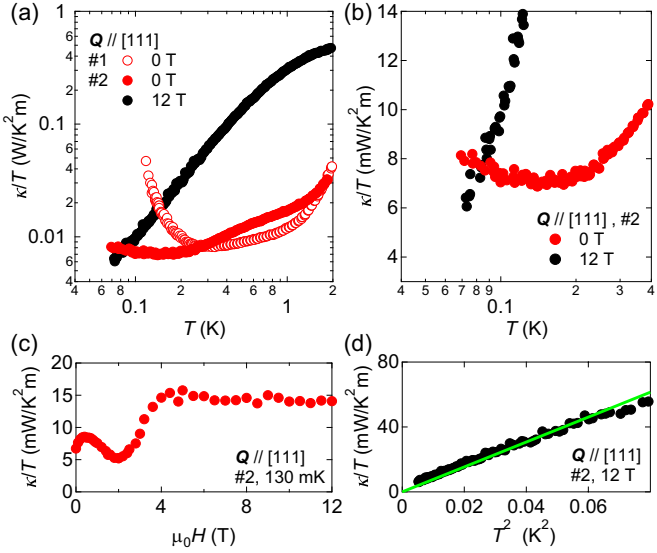}
		\caption{  
		(a)  Temperature dependence of $\kappa/T$  in zero field and at 12\,T of two different crystals  for $\bm{Q}$$\parallel$$[111]$.  The magnetic field is applied along [111] direction.   (b) $\kappa/T$ at low temperatures in zero field and at 12\,T. (c) Field dependence of $\kappa/T$ at 130\,mK.  (d)  $\kappa(12\,{\rm T})/T$, which corresponds to purely phononic conductivity $\kappa_{ph}/T$, plotted as a function of $T^2$. Green line represents the calculated $\kappa_{ph}/T$ from the specific heat of La$_2$Zr$_2$O$_7$ and effective size of the crystal.
		}
	\end{center}
	\vspace{-5mm}
\end{figure}

 The purely phononic thermal conductivity is confirmed by the $T$-dependence of $\kappa/T$ above saturation field. As shown in Fig.\,3(d), $\kappa/T$ at 12\,T increases in proportion to $T^2$ in regime-I.  This $T$-dependence is consistent with phonon conductivity in the limit of boundary scattering, where the phonon mean free path is comparable to the crystal size, 
\begin{equation}
\kappa_{ph}=\frac{1}{3}\beta v_s\ell_{ph}T^3,
\end{equation}
where $\beta, v_s$ and $\ell_{ph}$ are  the specific heat coefficient, the sound velocity and phonon mean free path, respectively.  According to Kimura {\it et al.}~\cite{Kimura}, $\beta$ is reported to be 2.5\,J/m$^3$K$^4$. From the Debye temperature $T_D=452$\,K  obtained from $\beta$, we find  $v_s=3700$\,m/s, using the relation $v_s=\frac{k_B}{\hbar}T_D(6\pi^2N/V)^{-1/3}$, where $N=88$ is the number of atoms per unit cell and $V$ is the unit cell volume. In the boundary scattering limit, $\ell_{ph}$ is given by the effective diameter $d_{eff}=2\sqrt{wt/\pi}$, where $w$ and $t$ are width and thickness of the crystal. Thus $\kappa_{ph}=0.77T^3$\,W/Km is obtained, which is shown by the green line in Fig.\,3d. Although this simple calculation of $\kappa_{ph}$ should be scrutinized, the calculation well reproduces the data. Moreover, $\ell_{ph}=3\kappa_{ph}/\beta T^3v_s$ becomes less than $d_{eff}$ above $\sim$0.4\,K, which is comparable to the temperature where $\kappa/T$ at 12\,T deviates from $T^2$ dependence (Fig.\,3d).  

It should be stressed that the anomalous enhancement of $\kappa/T$ in regime-I in zero field is not caused by phonon because of the following reasons. As shown in Fig.\,3(a) and (b), $\kappa/T$ in zero field exceeds $\kappa_{ph}/T[=\kappa(12\,{\rm T})/T]$ at very low temperatures.  Since phonon conduction is  in the boundary scattering limit,  $\kappa(12\,{\rm T})/T$ gives the upper limit of phonon conduction.   Therefore the results provide direct evidence that observed steep enhancement of $\kappa/T$ comes from spin excitations. In the present study, we focus on $\kappa/T$ in zero and high field regimes, because  in the intermediate field regime where Zeeman energy becomes comparable to the phonon  energy, a resonance between Zeeman gap and phonon  makes the analysis of the thermal conductivity complicated \cite{Berman}.

\begin{figure}[t]
	\begin{center}
		\includegraphics[width=1.0\linewidth]{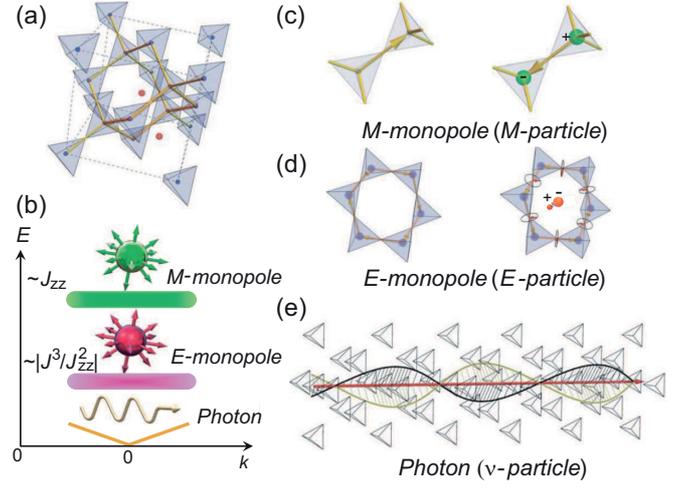}
		\caption{
		(a) Pyrochlore lattice structure with local $[111]$ spin axes (yellow thin and thick bonds). The blue (red and orange) dots form a diamond (dual-diamond) lattice.  (b) Magnetic excitations in a pyrochlore quantum spin liquid with spin ice correlations.  Three dispersive collective excitations are emergent; magnetic monopole ($M$-particle), electric monopole ($E$-particle) and photon ($\nu$-particle).  The $M$- and $E$-particles are gapped, while $\nu$-particls are gapless and their energy dispersion is linear in small momentum $k$. (c)  $M$-particle. The pair creation of an $M$-particle and its partner (green and magenta dots) from the left figure describe a spin-flip excitation along the local spin axis, which costs 	$\sim J_{zz}$, in the right figure. (d)  $E$-particle. The pair creation of an $E$-particle and its partner (red and orange dots) from the left figure describes a collective excitation in the right figure from the transverse components of the spins (red arrows in circles)  on a hexagon (thick line in (a)), which costs $\sim \frac{3 J_{\perp}^3 }{2 J_{zz}^2}$. (e) $\nu$-particle has energy dispersion relation, $E(k) =\hbar v_{p} |k|$, and mediates dynamic interaction between $E$ and $M$-particles. }
	\end{center}
	\vspace{-5mm}
\end{figure}

Our thermal conductivity in Pr$_2$Zr$_2$O$_7$ can be further understood by considering  a minimal pseudo-spin ($S$=1/2) Hamiltonian, 
  \begin{equation}
\mathcal{H} = \sum_{\langle i,j\rangle}\left[
J_{zz}S^z_i S^z_j+J_{\perp} (S^x_i S^x_j+S^y_i S^y_j)
\right],
  \end{equation}
where $J_{zz}$ and $J_{\perp}$ are the exchange couplings between the nearest neighbor spins in pyrochlore lattice [Fig.\,4(a)] along a local easy (Ising) axis and its transverse ($xy$) plane, respectively \cite{Hermele-PRB04,Benton-PRB12,Savary-PRL12,Ross-PRX11,Shannon-PRL12,Kato-PRL15, Machida}. Although additional terms such as $S_i^yS_j^x$ may be present, the essential features  of the present system can be  described by this minimal Hamiltonian. 
It is well understood that the Hamiltonian hosts three emergent excitations, i.e. magnetic monopole ($M$-particle), electric monopole ($E$-particle) and photon ($\nu$-particle) [Figs.\,4(b)-4(e)].  
A pair creation of $M$-particles describes a spin-flip excitation along the local spin axis, which costs $\sim J_{zz}$. The $M$-particles are present in classical spin-ice systems and have been detected by many experiments, while $E$- and $\nu$-particles 
have never been reported so far. 
A pair creation of $E$-particles generates an emergent electric field $\mathcal{E}$ along a bond direction of a dual diamond lattice, which corresponds to a pseudo-spin product around a hexagon [Fig.\,4(d)], $S_1^{+} S_2^{-} S_3^{+} S_4^{-} S_5^{+} S_6^{-} \propto \cos(\mathcal{E})$ \cite{Hermele-PRB04,Benton-PRB12}. Since the transverse components of the pseudo-spin describe local quadrupole moments in Pr$_2$Zr$_2$O$_7$ due to its non-Kramers' doublet, the pair creation of $E$-particles  describes collective excitations of quadrupole moments with an excitation gap $\sim3J_{\perp}^3/2J_{zz}^2$. 
Contrary to these gapped monopole excitations, $\nu$-particles are gapless collective excitations. 

We estimate the exchange couplings by fitting specific heat $C$ data of the same sample used for thermal conductivity with numerical calculation based on the full exact diagonalization to the pyrochlore 16-site cube cluster with periodic boundary condition. Our specific heat data, shown in Fig.~5, well reproduce the recently reported one of high-quality single crystals~\cite{Kimura14}. Figures \,5(a)-(c) show the comparison around $J_{\perp}/J_{zz}$=0.4, which describes best the experimental $C$($T$). From the fitting, we obtain $J_{\perp}/J_{zz}$=0.4 and $J_{zz}$=6.8\,K. Here, it should be noted that because the spin representation with $S^z=\pm1/2$ is used, $J_{zz}$ needs to be factored by 1/4 to compare with the value of coupling using $\sigma^z=\pm 1$. We mention that $J_{\perp}$ produces the high-temperature tail of $C$($T$), while the classical model with $J_{\perp}$=0 exhibits a much sharper peak in $C$($T$), which does not describe the experimental $C$($T$). 

\begin{figure}[t]
	\begin{center}
		\includegraphics[width=1\linewidth]{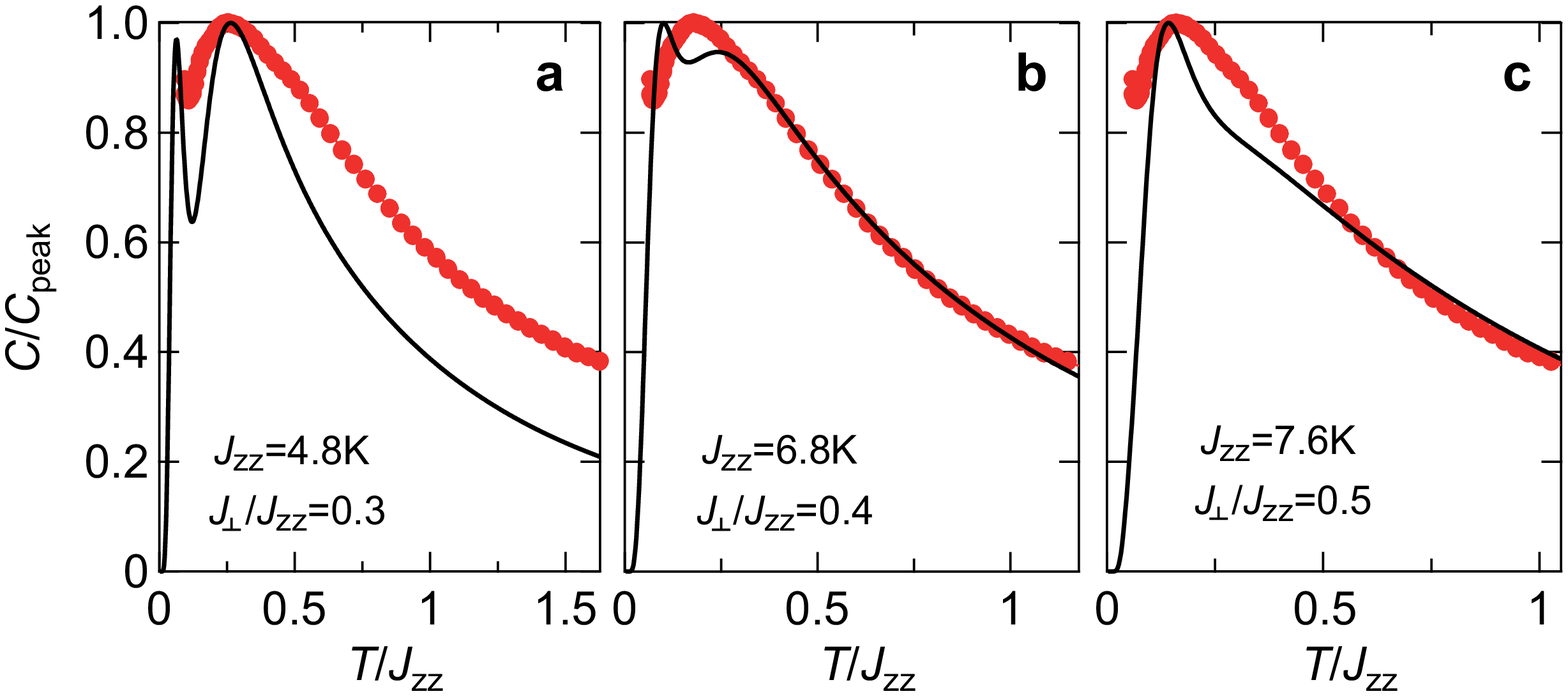}
	\end{center}
	\caption{{\bf Specific heat $C$ of the same sample piece used for thermal conductivity measurement.}  The normalized $C$ by the largest value $C_{\rm peak}$ is plotted against the normalized temperature by Ising exchange coupling $J_{zz}$. Red symbols and black line are experimental data and theoretical calculation, respectively. Contributions from nuclear, lattice and excited states of crystalline electric field levels~\cite{Kimura} are subtracted. The nuclear contribution is estimated by fitting the low-temperature specific heat below $T$=0.6\,K by 1/$T^2$. The lattice contribution is approximated by the specific heat of the non-magnetic analog La$_2$Zr$_2$O$_7$~\cite{Kimura}. The three results with different $J_{\perp}/J_{zz}$ around the best fit are shown. The numerical result describes best the experimental $C$($T$), when $J_{\perp}/J_{zz}$=0.4 and $J_{zz}$=6.8\,K.} 
\end{figure}

The peak of $\kappa/T$ in zero field at 70\,mK, which is far below  $J_{zz}$, is most naturally ascribed to the lowest-energy excitation in the QSL. 
Moreover, $\kappa(T) \propto 1/T$ for 70\,mK $\leq$ $T$ $\leq $ 150\,mK [Fig.\,1(c)].  Such a $T$-dependence is similar to that for phonons at high temperatures, which further supports gapless nature of excitations in regime-I. 
We estimate photon velocity as $v_p\sim{\frac{J_{\perp}^3}{J_{zz}^2} } \frac{a_0}{\hbar}$=90 ms$^{-1}$, which is in a similar order of magnitude to the previous theoretical estimations~\cite{Benton-PRB12, Savary-PRL12} ($a_0$ is the lattice constant). The specific heat of photons can be also estimated as $C_p \sim 1 \times 10^{-2}$ J  (mol)$^{-1}$ K$^{-1}$ at $T\sim$100\,mK~\cite{Benton-PRB12}, similar to theoretical estimations~\cite{Benton-PRB12, Savary-PRL12}.  Then, the photon mean free path  is obtained as $ l_p=3 \kappa_{p} /( C_p v_{p})\sim 2 $ $\mu$m.  Although our simple estimation should be scrutinized, the nearly ballistic propagation of photons seems to be realized at low $T$. 

\begin{figure}[t]
	\begin{center}
		\includegraphics[width=0.7\linewidth]{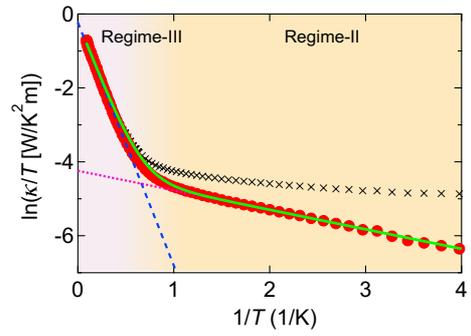}
		\caption{
		Arrhenius plot of $\kappa/T(T)$ before (black crosses) and after subtraction of the $1/T^2$-term (red circles) found in regime-I [(a)]. Blue dashed and pink dotted lines are the exponential fits in regime III and II, and green solid line is a two-gap fit with $\Delta_{\rm III}=6$\,K and $\Delta_{\rm II}=0.53$\,K. 
		}
	\end{center}
	\vspace{-5mm}
\end{figure}

Above 1.5\,K in regime-III,  $\kappa/T$ exhibits a steep increase with $T$ (Fig.\,1(a)).   As shown in Fig.\,6,  $\kappa/T$ increases with a thermal-activation behavior, $\kappa(T)/T\propto  \exp(-\Delta_{{\rm III}}/T)$.  The estimated energy gap $\Delta_{{\rm III}}\approx 6$\,K is comparable to $J_{zz}$, which is consistent to the $M$-particles excitations.  We note that the thermally activated behavior has been reported by magnetic susceptibility and specific heat measurements, and has been attributed to the $M$-particle excitations \cite{Kimura}. 

We next discuss the $T$-dependence of $\kappa/T$ in regime II, where a hump structure is seen (Fig.\,1(b)).  With a subtraction of the $1/T^2$-dependence in regime-I, $\kappa/T$ in regime-II can be fitted to the exponential dependence  $\exp(-\Delta_{\rm II}/T)$ with $\Delta_{\rm II}=0.53$\,K, as shown in Fig.\,6.  The amplitude of  $\Delta_{\rm II}$ is close to the excitation energy of $E$-particles, $\sim3J_{\perp}^3/2J_{zz}^2$$\sim$0.6\,K. Thus  it is tempting to attribute the hump structure  in regime-II to $E$-particles.

We remark that the minimal model with the bilinear exchange ($J_{\perp}$) term is not the only way to describe the QSL. In recent literature, an alternative way for the QSL is proposed. Namely, the non-Kramers Pr ions feel the linear transverse-field terms from random impurities in Pr$_2$Zr$_2$O$_7$ \cite{Wen2017, Savary2017}.  Both of the linear and bi-linear terms could exist in reality, and the cleaner samples would have the stronger bilinear couplings. The QSL can be even stabilized under the presence of impurities.  

Finally, we comment on the field dependence of $\kappa/T$ in the regime-I. 
The initial steep increase of $\kappa(H)/T$  in regime-I [Figs.\,2(a) and 3(c)], where no previous experiments have reached in other pyrochlore materials, indicates that the heat carriers in this regime are essentially different from those of the regimes II and III.  This field-induced enhancement of $\kappa(H)/T$ should come from reduced scattering between photons and monopoles under magnetic fields.  Thus, $\kappa(H)/T$ appears to be consistent with the $U(1)$ quantum spin liquid with gapless $\nu$-particles below 70\,mK realized in Pr$_2$Zr$_2$O$_7$.


\section{Summary}
We have investigated the low-energy itinerant quasiparticle excitations in the quantum spin-liquid candidate Pr$_2$Zr$_2$O$_7$ by thermal transport measurements. We find that the thermal conductivity shows a dramatic enhancement below $\sim200$\,mK, which can be attributed to emergent photons ($\nu$-particle), coherent gapless spin excitations in a spin-ice manifold. The two gap behavior observed at higher temperatures is consistent with two different monopoles, i.e. magnetic ($M$-particle) and electric monopoles ($E$-particle). The present observation  implies that fractionalized quasiparticles and gauge field fluctuations play fundamental roles in quantum magnets with three dimensional QSL ground states.

\begin{acknowledgment}

We thank  L. Balents, G. Chen, M.J.P. Gingras, S. Onoda, K. Totsuka, C. Broholm  for useful discussions. This work was supported by Grants-in-Aid for Scientific Research (KAKENHI) (No.\,25220710, No.\,26800199, No.\,15H02106, No.\,15K13533, 15K13521, No.\,16H02206, No.\,16H02209), Grants-in-Aid for Scientific Research on Innovative Areas ``Topological Materials Science" (No.\,15H05852)  and ``J-Physics" (15H05882, 15H05883), and Program for Advancing Strategic International Networks to Accelerate the Circulation of Talented Researchers (No.\,R2604) from Japan Society for the Promotion of Science (JSPS) and by CREST, Japan Science and Technology Agency. E.-G. M. acknowledges the financial supports from the POSCO Science Fellowship of POSCO TJ Park Foundation and NRF of Korea under Grant No. 2017R1C1B2009176. 

\end{acknowledgment}


\begin{thebibliography}{9}
\bibitem{Wen2002}
X. G. Wen, 
Phys. Rev. B {\bf 65}, 165113 (2002).
	
\bibitem{Sachdev}
S. Sachdev, 
Nat. Phys {\bf 4}, 173 
(2008).
	
\bibitem{Balents}
L. Balents, 
Nature {\bf 464}, 199 
(2010).

	


	
	




\bibitem{Castelnovo-Nature08}
C. Castelnovo, R. Moessner and S. Sondhi, 
Nature {\bf 451}, 42-45 (2008).

\bibitem{Jaubert} 
L. D. C. Jaubert and P. C. W. Holdsworth, 
Nat. Phys {\bf 5} 258 (2009).



\bibitem{Morris-Science09}
D. J. P. Morris {\it et al.}, 
Science {\bf 326}, 411 
(2009).

\bibitem{Ramirez-Nature99}
A. P. Ramirez, A. Hayashi, R. J. Cava, R. Siddarthan and B. S. Shastry, 
Nature {\bf 399}, 333-335 (1999).

\bibitem{Bramwell-Science01}
S. T. Bramwell and M. J. P. Gingras, 
Science {\bf 294}, 1495 
(2001).

\bibitem{Bramwell-PRL01}
S. T. Bramwell {\it et al.}, 
Phys. Rev. Lett. {\bf 87}, 047205 (2001).

\bibitem{Fennell-Science09}
T. Fennell {\it et al.}, 
Science {\bf 326}, 415 
(2009).

%
%


\bibitem{Hermele-PRB04}
M. Hermele, M. P. A. Fisher, L. Balents,
Phys. Rev. B {\bf 69}, 064404 (2004).

\bibitem{Benton-PRB12}
O. Benton, O. Sikora and N. Shannon, 
Phys. Rev. B {\bf 86}, 075154 (2012).

\bibitem{Onoda}
S. Onoda and Y. Tanaka, 
Phys. Rev. Lett. {\bf 105}, 047201 (2010).

\bibitem{Ross-PRX11}
K. A. Ross, L. Savary, B. D. Gaulin and L. Balents, 
Phys. Rev. X {\bf 1}, 021002 (2011).

\bibitem{Savary-PRL12}
L. Savary and L. Balents, 
Phys. Rev. Lett. {\bf 108}, 037202 (2012).

\bibitem{Shannon-PRL12}
N. Shannon, O. Sikora, F. Pollmann, K. Penc and P. Fulde, 
Phys. Rev. Lett. {\bf 108}, 067204 (2012).

\bibitem{Gingras-RPP14}
M. J. P. Gingras and P. A. McClarty, 
Rep. Prog. Phys. {\bf 77}, 056501 (2014).

\bibitem{Kato-PRL15}
Y. Kato and S. Onoda, 
Phys. Rev. Lett {\bf 115}, 077202 (2015).

\bibitem{Machida} 
Y. Machida, S. Nakatsuji, S. Onoda, T. Tayama and T. Sakakibara, 
Nature {\bf 463}, 210 (2010).





\bibitem{Kimura} 
K. Kimura {\it et al.}, 
Nat. Commun. {\bf 4}, 1934 (2013). 

\bibitem{Matsuhira} 
Matsuhira {\it et al}., J. Phys. Conf. Ser. {\bf 145}, 012031 (2009)

\bibitem{Kimura14} 
K. Kimura and S. Nakatsuji, 
JPS Conf. Proc. {\bf 3}, 014027 (2014).

\bibitem{Petit} 
Petit. {\it et al}., Phys. Rev. B {\bf 94}, 165153 (2016)

\bibitem{Chang} 
L.-J. Chang, S. Onoda, Y. Su, Y.-J. Kao, K.-D. Tsuei, Y. Yasui, K. Kakurai and M. R. Lees, 
Nat. Commun. {\bf 3}, 992 (2012).

\bibitem{Tokiwa} 
Y. Tokiwa {\it et al.}, 
Nat. Commun. {\bf 7}, 10807 (2016).


\bibitem{Gardner-RMP10} 
J. S. Gardner, M. J. P. Gingras and J. E. Greedan, 
Rev. Mod. Phys. {\bf 82}, 53 (2010). 

\bibitem{Sondhi12}
C. Castelnovo, R. Moessner and S. L. Sondhi, 
Ann. Rev. Con. Matt. Phys. {\bf 3}, 35 (2012).

%
\bibitem{SaitoPRE96}
K. Saito, S. Takesue and S. Miyashita, 
Phys. Rev. E  {\bf 54}, 2404 (1996).
%


\bibitem{Sologubenko-PRB01}
A. V. Sologubenko, K. Giann$\rm{\grave{o}}$, H. R. Ott, A. Vietkine and A. Revcolevschi, 
Phys. Rev. B {\bf 64}, 054412 (2001).


\bibitem{Hess-PRL04}
C. Hess, H. ElHaes, B. B$\rm{\ddot{u}}$chner, U. Ammerahl, M. H$\rm{\ddot{u}}$cker and A. Revcolevschi, 
Phys. Rev. Lett. {\bf 93},  027005 (2004).


\bibitem{Yamashita-NatP}
M. Yamashita {\it et al.}, 
Nat. Phys {\bf 5}, 44 
(2009).

\bibitem{Yamashita-Science}
M. Yamashita {\it et al.}, 
Science {\bf 328}, 1246 
(2010).

\bibitem{future}
K. Kimura, {\it et al.},  
To be published elsewhere.

\bibitem{Koohpayeh}
S.M. Koohpayeh, J.-J. Wen, B.A. Trump, C.L. Broholm, T.M. McQueen,
J. Cryst. Gr. {\bf 402}, 291. (2014)


\bibitem{Toews} 
W. H. Toews, S. S. Zhang, K. A. Ross, H. A. Dabkowska, B. D. Gaulin and R. W. Hill, 
Phys. Rev. Lett. {\bf 110}, 217209 (2013).

\bibitem{Berman} 
R. Berman, 
{\it Thermal Conduction in Solids} (Oxford Univ Press, Oxford, 1976). 

\bibitem{Wen2017}
J.-J. Wen {\it et al.}, 
Phys. Rev. Lett. {\bf 118}, 107206 (2017).

\bibitem{Savary2017}
L. Savary and L. Balents, 
Phys. Rev. Lett. {\bf 118}, 087203 (2017).










\end{thebibliography}
\end{document}